\documentclass[preprint,prl,superscriptaddress,showpacs,byrevtex]{revtex4}
\usepackage{mathptmx,amssymb}
\usepackage{amsmath}
\usepackage{graphicx}
\usepackage{color}
\usepackage{epstopdf}

\begin{document}

\title{Optimum reduction of the dynamo threshold by a ferromagnetic layer located in 
the flow}

\author{J. Herault, F. P\'etr\'elis}
\affiliation{Laboratoire de Physique Statistique, Ecole Normale
Sup\'erieure, CNRS, Universit\'e P. et M. Curie, Universit\'e Paris
Diderot, Paris, France}

\begin{abstract}
We consider a fluid  dynamo model generated by the flow on both sides of a moving layer. The  magnetic permeability of the layer is larger than that of the flow. We show that there exists an optimum value of  magnetic permeability for which the critical magnetic Reynolds number for dynamo onset is smaller than for  a nonmagnetic material and also smaller than for a layer of infinite magnetic permeability. We present a  mechanism    that provides an explanation for  recent experimental results. A similar effect occurs when the   electrical conductivity of the layer is large. 

\end{abstract}
\maketitle



Several properties of hydrodynamic instabilities 
are  highly sensitive to the boundary conditions of the system. This is
generically the case for the   instability threshold. Changing the
boundary conditions modifies the form of the unstable mode which 
results in a change in the efficiency of the processes responsible for the
instability. This, in turn, modifies the values of the system parameters that are necessary to reach  the   instability threshold.  
In the canonical example of Rayleigh-B\'enard convection in a layer of infinite horizontal extension with fixed temperature at the top and bottom boundaries,  the threshold, measured by  the Rayleigh number, differs by a factor close to three depending if the boundary conditions for the velocity
field are  no-slip or stress-free  \cite{chandra}.

For  the dynamo instability,  the boundary conditions for the magnetic field 
are also crucial. 
For astrophysical applications, the physical properties  of the external
or internal boundaries are sometimes so poorly constrained that models are free 
to investigate  boundaries with completely different
properties.  
In experiments,   dynamos are difficult to observe because they are generated
by turbulent flows of liquid metals  driven  at very large velocities.
Experiments  are often operated at the limit of the achievable parameters. Therefore  even a small reduction of   the dynamo threshold is always
welcome while an increase can easily turn a successful experiment into a
failure.

The Von Karman Sodium experiment (VKS) \cite{P1} consists of 
a swirling flow of liquid sodium that is driven  by the
counter-rotation of two disks fitted with blades.  A strong shear is present in the mid-plane  between the disks. This shear converts poloidal to toroidal magnetic field through the $\omega$-effect. Moreover, the flow is helical close to the disks, which generates an $\alpha$-effect that can  convert toroidal to poloidal magnetic field \cite{gafd}. 
A dynamo is only observed when the propellers are made
of soft iron which has large magnetic permeability; thus the boundary conditions are also of great importance. The effects of boundary conditions on the VKS dynamo have been
investigated in several studies.
The  boundaries of the container were initially considered \cite{ravelet}.
Experimentally,  the nature of these boundaries was observed to be far
less important that the nature of the disks.
The first attempt to model the disks was to assume  that their large magnetic permeability 
constrains  the  field to be   purely normal
to the propeller boundaries \cite{gafd}, \cite{nore}, \cite{gissinger}. 
A second idea relies on treating the effect of  the blades as  an inhomogeneous 
magnetic permeability  in the azimuthal direction. Inhomogeneities of
the electrical conductivity are known to be able to generate a dynamo
provided that an $\omega$-effect acts close to the boundary \cite{wicht}. A
similar idea was put forward to explain numerical simulations of the VKS
experiment \cite{guisecke}. The mechanism has been identified and calculated analytically
in model geometries \cite{gallet1}, \cite{gallet2}. From these
calculations, the effect of inhomogeneities appears as an additional induction mechanism that converts toroidal  to poloidal field. However, the  
effect is not very efficient and very large magnetic Reynolds numbers are
necessary to generate a dynamo.
It has also been suggested that the motion of the soft iron disk leads to
an increase in the  $\omega$-effect and that the effect depends on the magnetic
Reynolds number based on the disk properties \cite{verhille}.

Assuming that a boundary can be modeled as a medium of large magnetic
permeability, one can classify the possible effects  into two groups. (i)  
The change in boundary  conditions lowers the onset of a mode that would be a dynamo
at higher onset with nonmagnetic material. (ii) The  boundary
provides a new  amplification mechanism that is responsible for the dynamo
action and without which a similar dynamo would not exist.
In the mechanisms reported so far, the effect of the blades (inhomogenous
magnetic permeability) belongs to group (ii) while the other effects
are of group (i).

We present here another  mechanism that belongs to  group (i). We focus on a single disk and  consider the structure of the magnetic field on both sides of
the disk. Indeed, the symmetries  of the flow constrain several 
properties of the  induction process. For instance, shears of opposite signs are located on either side of the propeller. Then, we show that a large but  finite value of the magnetic permeability of the disk results in 
a decrease of the dynamo threshold compared to the case of a nonmagnetic 
or ideal ferromagnetic material. 

A simple geometrical configuration is considered to model the disk and the
flow. The configuration  is presented first, along with the methods used  to solve the
problem. Numerical results are then presented and extended to the case of a change in the electrical conductivity. We finish with an explanation of the  increased
efficiency of the dynamo for certain optimal values of the disk physical
properties.


{\bf Presentation of the calculation}

\begin{figure}[htb!]
\begin{center}
\includegraphics[width=120mm]{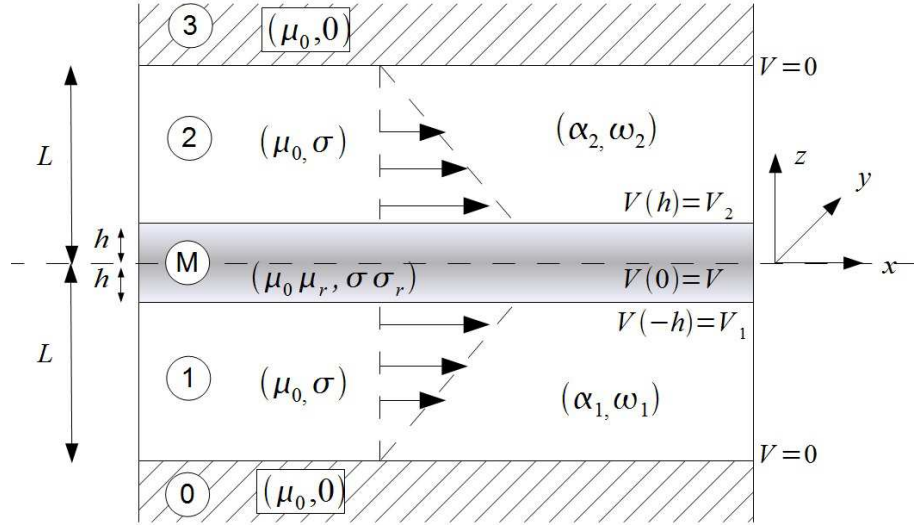}
\caption{Sketch of the model. A layer $(M)$ moves at constant velocity $V$ and is located in-between two medium $(1)$ and $(2)$ in which $\omega$ and $\alpha$ effects are present. The outer medium $(0)$ and $(3)$ are insulating.
}
\label{fig0}
\end{center}
\end{figure}

To simplify the problem, we consider a cartesian geometry  as sketched in fig. \ref{fig0}.  The $x, y$ and $z$ coordinates correspond to the $\theta, r$ and $z$ coordinates in cylindrical geometry.  A fluid
of magnetic permeability $\mu_0$ and electrical conductivity $\sigma$ is
located between $z=-L$ and $z=-h$ and between $z=h$ and $z=L$.  A third layer of magnetic permeability $\mu_0 \mu_r$ and electrical conductivity $\sigma \sigma_r$ is located between these two
layers. The outer medium $|z| \ge L$ is  insulating and non-magnetic,  
as is a vacuum. 

The induction processes in the fluid layers are modeled using the equations of mean-field magnetohydrodynamics. We thus solve
\begin{equation}
\frac{\partial {\bf B}}{\partial t}=\nabla \times \left({\bf v} \times {\bf B} + \alpha {\bf B}\right)+\eta \nabla^2 {\bf B}\,, 
\label{eq0}
\end{equation}
where $\bf v$ is the large-scale velocity field and $\alpha$ is the tensor of the $\alpha-$effect. This tensor  describes the effect of small-scale velocity structures on the large-scale magnetic field   \cite{moffatt}. 
In the first (resp. second) layer, two induction processes are considered:
a shear of strength $\omega_1$ (resp. $\omega_2$) associated to a
velocity $ v^{(1)}_x=\omega_1 (z+L)$ (resp. $ v^{(2)}_x=\omega_2 (z-L)$) in the
$x$ direction and an $\alpha$-effect $\alpha^{(1)}$ (resp. $\alpha^{(2)}$).
The shear converts magnetic field in the $z$-direction into a field in the
$x$-direction. The $\alpha$-effect   converts magnetic field along $x$ in a
current along $x$, {\it i.e.} $\alpha^{(u)}_{i,j}=\alpha_u \delta_{i,x} \delta_{j,x}$. The mid-layer translates  at a uniform speed $V$.  
We use $L$ and $L^2\mu_0 \sigma$ as the units of length and  time, respectively and we define
the following dimensionless numbers $R_{\omega_i}=\mu_0 \sigma \omega_i
L^2$, $R_{\alpha_i}=\mu_0 \sigma \alpha_i L$, for $i=1,2$ and $R_V=\mu_0
\sigma V L$.

To solve   equation \ref{eq0}, we seek for modes of the form ${\bf b}(z)
\exp{i(pt +k y)}$ where $p$ is the growth rate and $k$ the wave-vector in
the $y$-direction. Our restriction to solutions independent of the $x$-coordinate  corresponds to  axisymmetric large-scale fields in the cylindrical geometry of the VKS experiment.   Using the divergence-free  property of the magnetic field, we can eliminate $b_y$ from
the problem and we have to solve the following coupled equations for the
$x$ and $z$ components. In layer $j$, for $j=1, 2$, equation \ref{eq0} reads 
\begin{equation}
\label{eqx2}
\left( p+k^2 \right) b_x ^{(j)}-  \partial_{zz} b_x^{(j)}-
R_{\omega_j} b_z^{(j)}=0\, ,
\end{equation}

\begin{equation}
\left( p+  k^2 \right) b_z ^{(j)}-  \partial_{zz} b_z^{(j)}
+R_{\alpha_j} i k  b_x^{(j)}=0 \, .
\label{eqz2}
\end{equation}
In the mid-layer $M$, we solve a diffusion equation with a relative dimensionless diffusivity $\eta_r=1/(\sigma_r \mu_r)$, 
\begin{equation}
\left( p/ \eta_r+ k^2  \right) \textbf b^{(M)}= \partial_{zz}\textbf b^{(M)}
\label{eqmid3}
\end{equation}
while in the outer insulating medium, we solve the equation $\nabla \times \bf B=0$. Using the divergence-free condition, this reduces 
to 
\begin{equation}
\partial_{zz} \textbf b_z  ^{(j)}=k^2 \textbf b_z ^{(j)}
\end{equation}
and $\partial_{y} \textbf b_x  ^{(j)}=\partial_{z} \textbf b_x  ^{(j)}=0$, for $j=0, 3$. 
These equations must be supplemented with boundary conditions. Using
continuity of the normal component of the magnetic field, of the
tangential component of ${\bf b}/\mu$, and of the tangential component of the
electric field, we obtain four relations at each interface  between medium $i$ and $j$

\begin{equation}
   b_z  ^{(i)}=b_z  ^{(j)} ,  \quad \quad
   \frac{b_x  ^{(i)}}{\mu_r^{(i)}   }=\frac{b_x  ^{(j)}}{\mu_r^{(j)} }  ,  \quad \quad
  \frac{\partial_z b_z  ^{(i)}}{\mu_r^{(i)}   }=\frac{\partial_z b_z  ^{(i)}}{\mu_r^{(j)} }  ,  \quad \quad E_y^{(i)}=E_y^{(j)}
\label{eqzBL1}
\end{equation}
where 
\begin{equation}
E_y^{(i)}= u_x^{(i)} b_z  ^{(i)}+\frac{1}{\mu_r^{(i)} \sigma_r^{(i)}} \partial_z b_x  ^{(i)}.
\label{eqzBL2}
\end{equation}

The eigenvalues $p$  of this linear problem are found by seeking 
solutions that
decrease to zero as  $|z|$ tends to infinity.  The solutions in the
outer medium involve two unknown constants and using the boundary
conditions, these constants  can be eliminated to obtain two new boundary conditions for the field in
medium $1$ (resp. $2$) at $z=-1$ (resp. $z=1$). These new boundary conditions do
not involve the value of the field in the outer medium.

We now have  to calculate the possible values of $p$ and the associated 
spatial structure of the magnetic field eigenmodes.
We have solved this problem with two different and  standard methods. The first method is to solve analytically  the equations in each layer. Each
solution in layer $1$, $M$ and $2$ involves four unknown constants as the
problem can be written as one fourth-order differential equation  for
a single function (either $b_x$ or $b_z$) in medium $1$ or $2$, and as two separated second-order equations for $b_x$ and $b_z$ in medium $M$. The twelve boundary conditions
form a linear set of
equations for the unknown constants. The existence of non-zero solutions
requires the  cancellation of the determinant of this set of equations. The
problem has thus been turned into that of finding zeros of a determinant. This equation selects the possible values of $p$.

The second method consists in solving the equations numerically using a
spatial discretization of $N$ gridpoints in each layer. The values of
the two components of the field at these $3 N$ gridpoints form a vector
of size $6 N$. We obtain a set of $6 N$ linear equations by writing the
two equations (Eq. (\ref{eqx2}) and (\ref{eqz2}) or the $x$ and $z$ component of Eq. (\ref{eqmid3})) evaluated at the $3 N$ gridpoints and by taking into account
the boundary conditions. Eigenvalues of the associated matrix are
possible  growth rates $p$.

Each of the two methods has its own advantages and drawbacks. For the
first method, one must find zeros of a complicated function. This defines an implicit expression of  $p$ as a function of the problem parameters.  Standard
methods require  initial guesses for the zeros and 
inaccurate guesses can result in solutions that are not the most
unstable. In contrast, with the second method,  approximations of  $6
N$ eigenvalues are found but these are only  approximations and one
needs to check the convergence of the numerical results when $N$ is
increased.
For the results presented below, we have checked that both methods give
the same results provided $N$ is large enough. Typically, with $N$ of order $100$, the difference between the two methods is smaller 
than  $0.05 \%$ for the onset of the first eigenmodes. 

The problem involves  a large number of parameters. In  dimensionless
form there are nine  parameters, namely
$k$, $h$, $\sigma$, $\mu$, $R_{\omega_1}$, $R\alpha_1$, $R_{\omega_2}$, $R\alpha_2$, $R_V$. Obviously one cannot consider performing a parametric study for such a
general problem.  What we have done is  to highlight one
particular effect observable in this configuration: the decrease of the
dynamo threshold when the mid-layer has a large but finite magnetic
permeability (or electrical conductivity). We have checked that the
effect is robust by changing drastically some hypotheses, aspect ratio or symmetry of
the flow. 

{\bf Results}

\begin{figure}[htb!]
\begin{center}
\includegraphics[width=70mm,height=70mm]{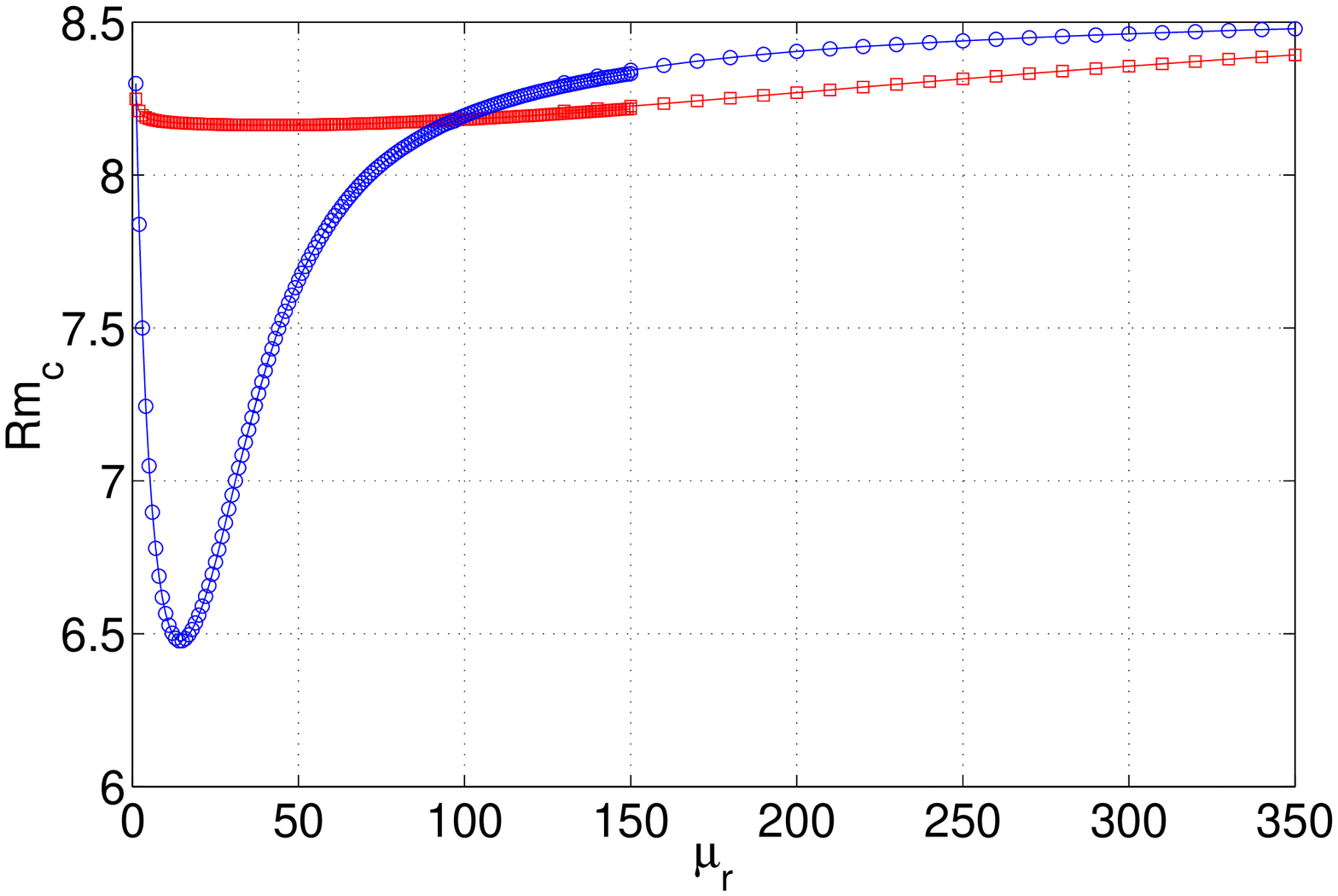}
\includegraphics[width=70mm,height=70mm]{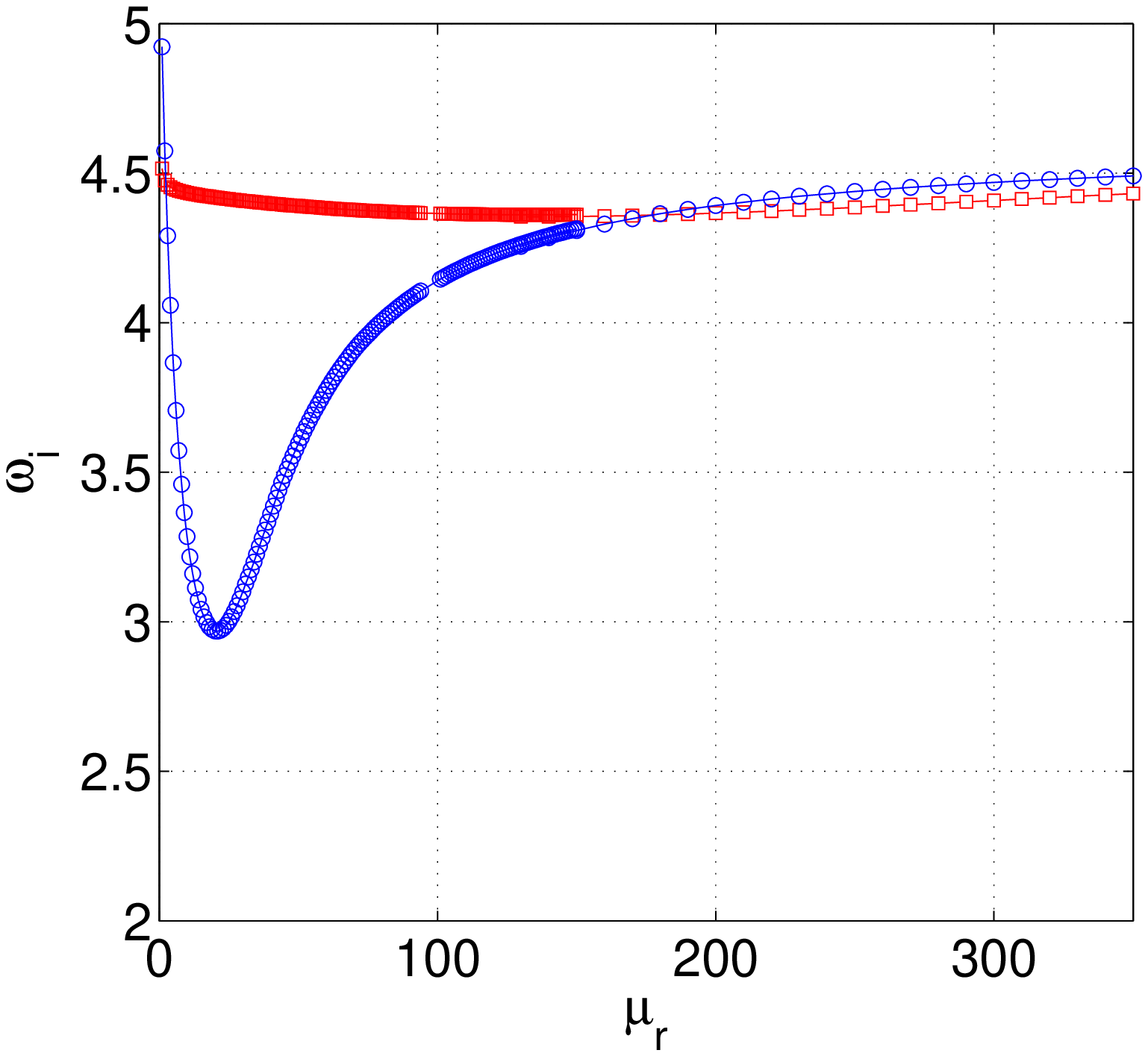}
\caption{Left: Onset of the two eigenmodes with largest growth rates as a
function of the magnetic permeability $\mu_r$  of the mid-layer; for a
symmetric configuration $R_{\omega_1}=R\alpha_1=-R_{\omega_2}=-R\alpha_2=R_V=$Rm and $\sigma_r=1$, $h=1/20$, $k=1$. Right: Frequencies of the two
eigenmodes with largest growth rate. Blue circles  (resp. red squares) are the even (resp.
odd) mode.
}
\label{fig1}
\end{center}
\end{figure}

We first present  a situation in which the induction mechanisms are of same intensity but have opposite signs 
in the two layers. We thus choose $R_{\omega_1}=R\alpha_1=-R_{\omega_2}=-R\alpha_2=$Rm.
We consider that the mid-layer is moving at the same speed as the flow at
$|z|=h$, thus $R_V=$Rm. The onsets Rm$_c$ of the two most unstable modes 
are presented in fig. \ref{fig1} for $k=1$ and $h=1/20$.   Both curves display a minimum  at
a finite value of $\mu_r$. For one of the modes, the
onset reduction
compared to $\mu_r=1$ is of the order of $25 \%$. For the other mode, the reduction is very small, of the order of  $1 \%$.  
We also note that both modes are oscillatory which amounts to  propagation
in the $y$-direction.
This is expected since the problem is not invariant under $y\rightarrow
-y$ (as the $\alpha$ coefficients are pseudo-scalars).

\begin{figure}[htb!]
\begin{center}
\includegraphics[width=70mm,height=70mm]{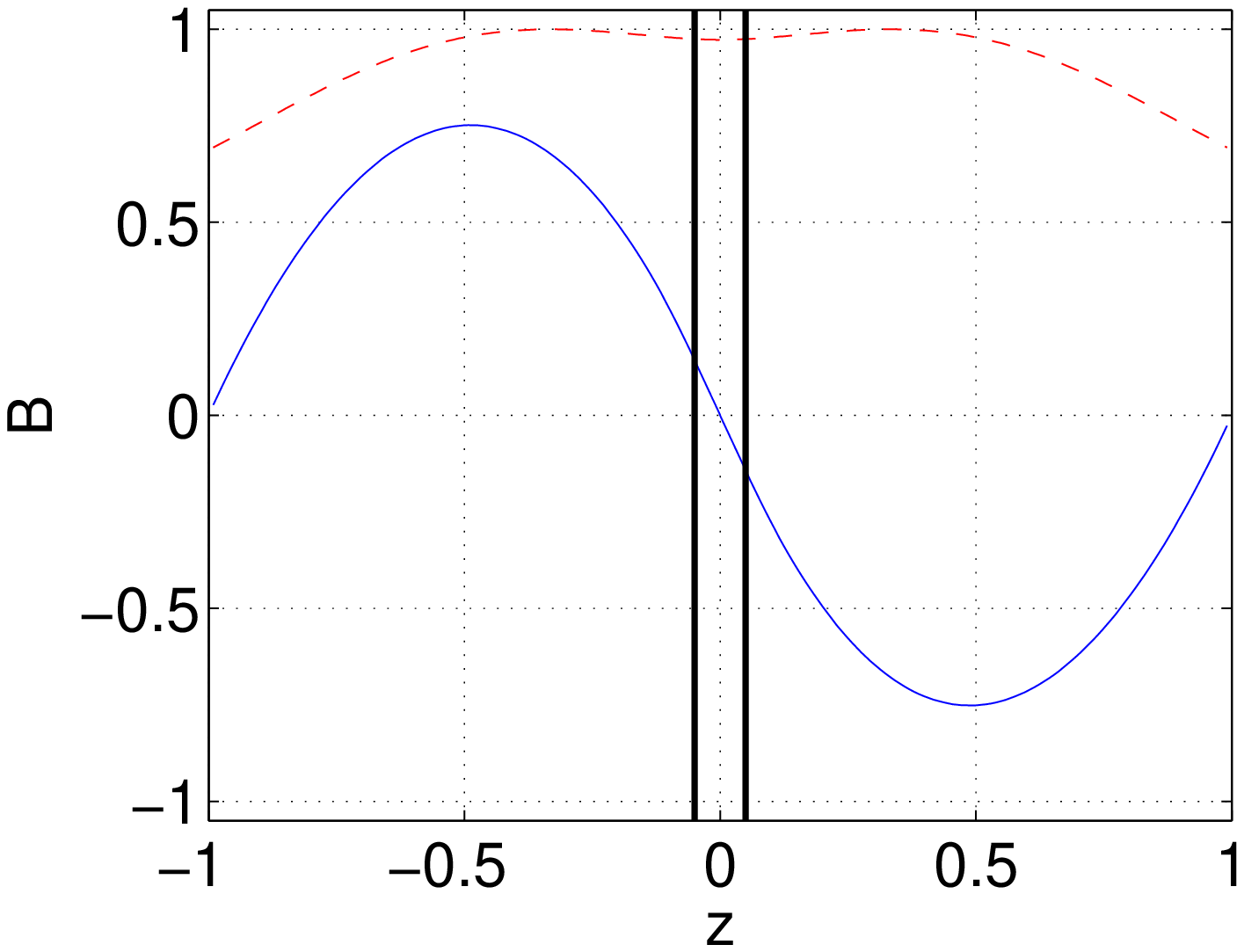}
\includegraphics[width=70mm,height=70mm]{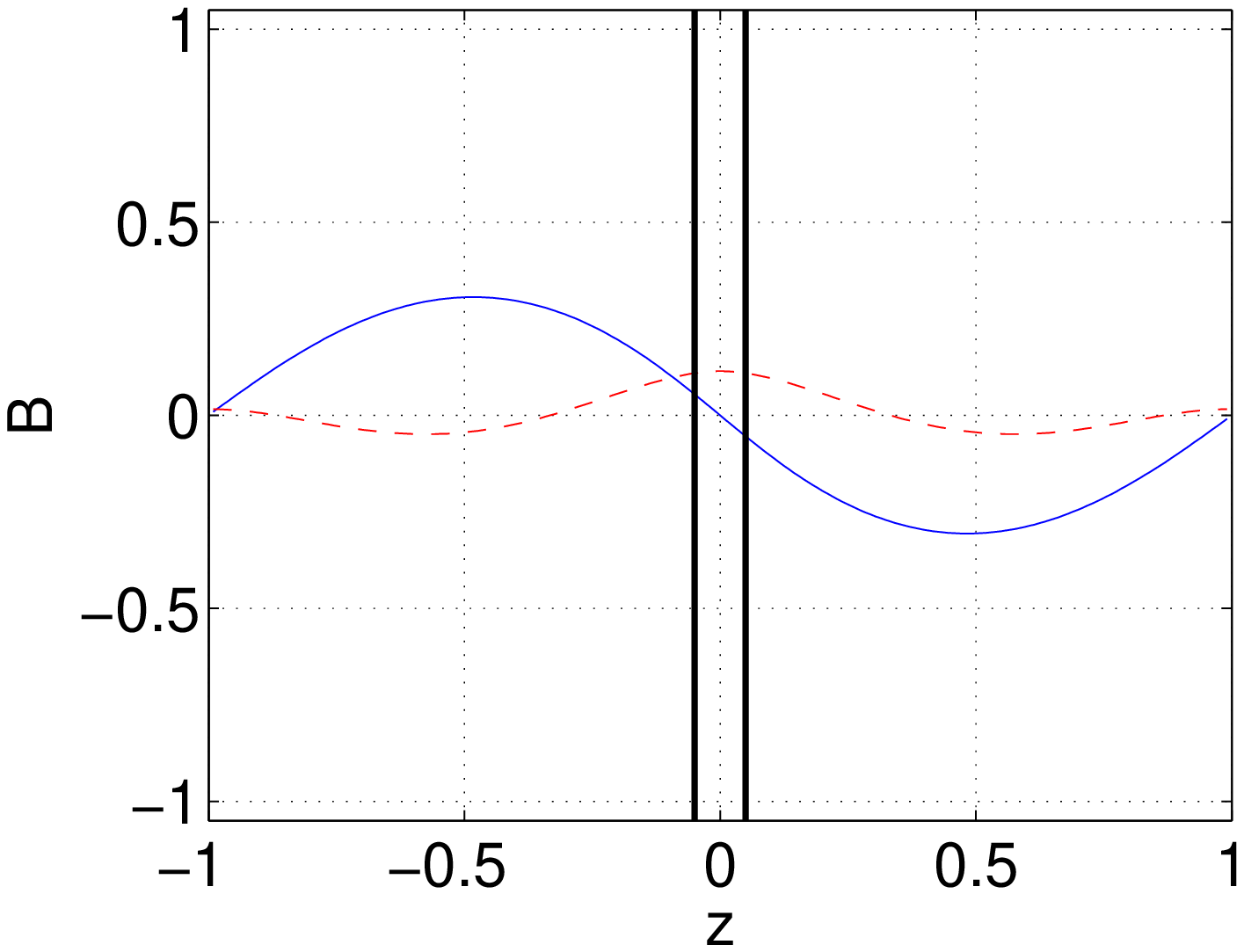}
\includegraphics[width=70mm,height=70mm]{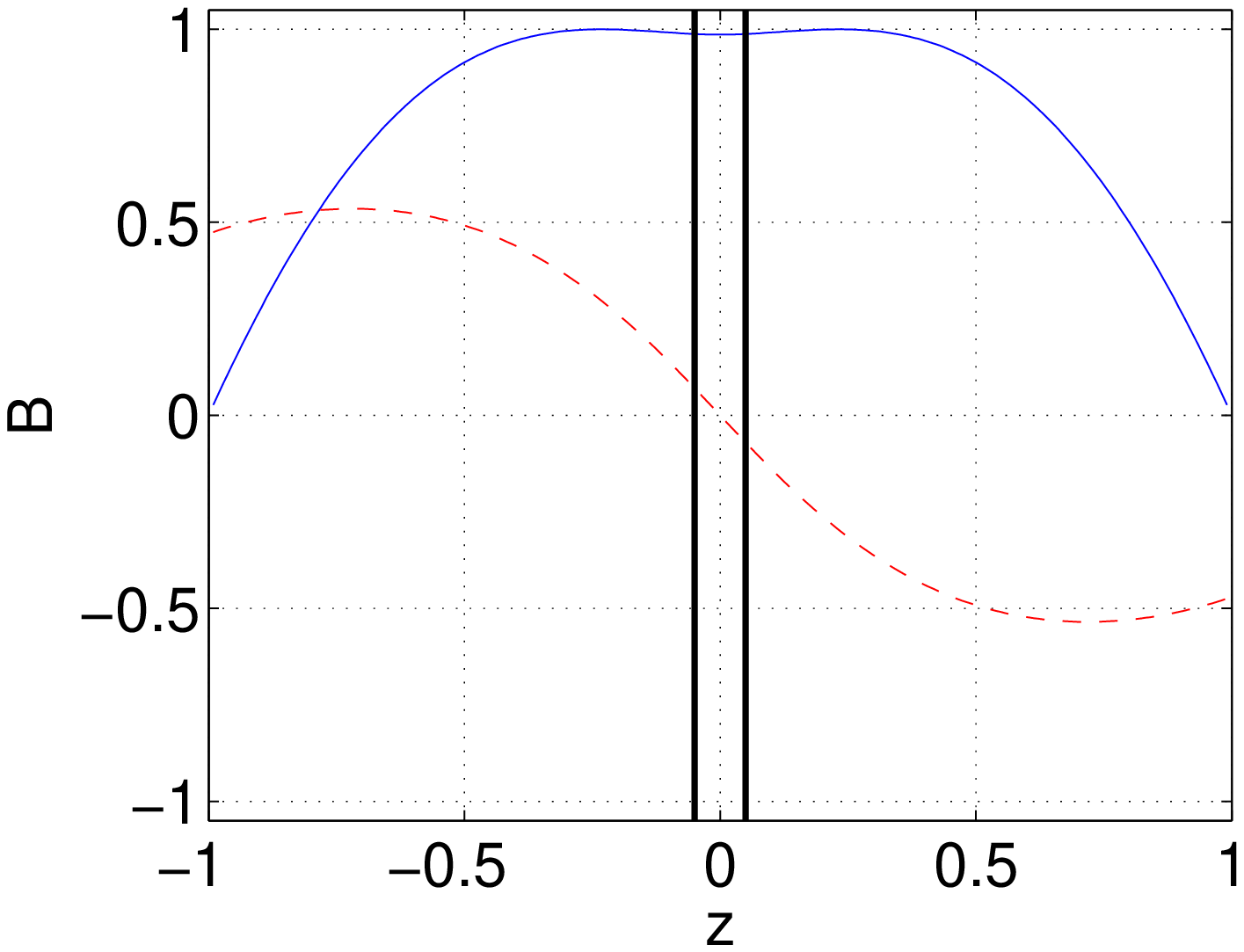}
\includegraphics[width=70mm,height=70mm]{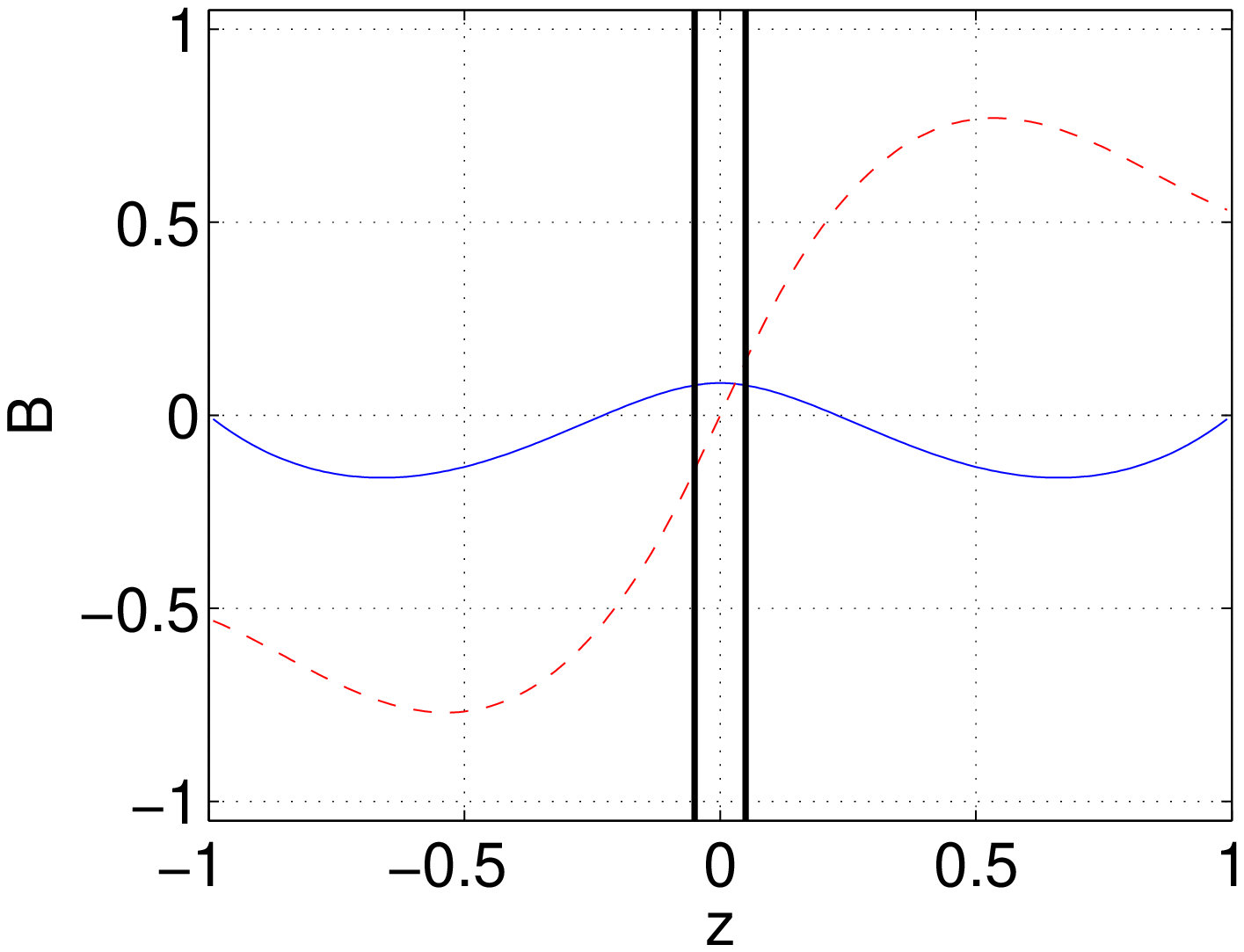}
\caption{Eigenmodes at their onset for a symmetric configuration and $\mu_r=1$, $\sigma_r=1$, top: most unstable mode  Rm$=8.25$. 
Bottom: second most unstable mode Rm$=8.3$. Real part (resp. imaginary part) are displayed on the left (resp. right). The $x$ (toroidal) component is the continuous (blue) curve. The $z$ (poloidal) component is the dashed (red) curve. 
}
\label{fig1a}
\end{center}
\end{figure}

\begin{figure}[htb!]
\begin{center}
\includegraphics[width=70mm,height=70mm]{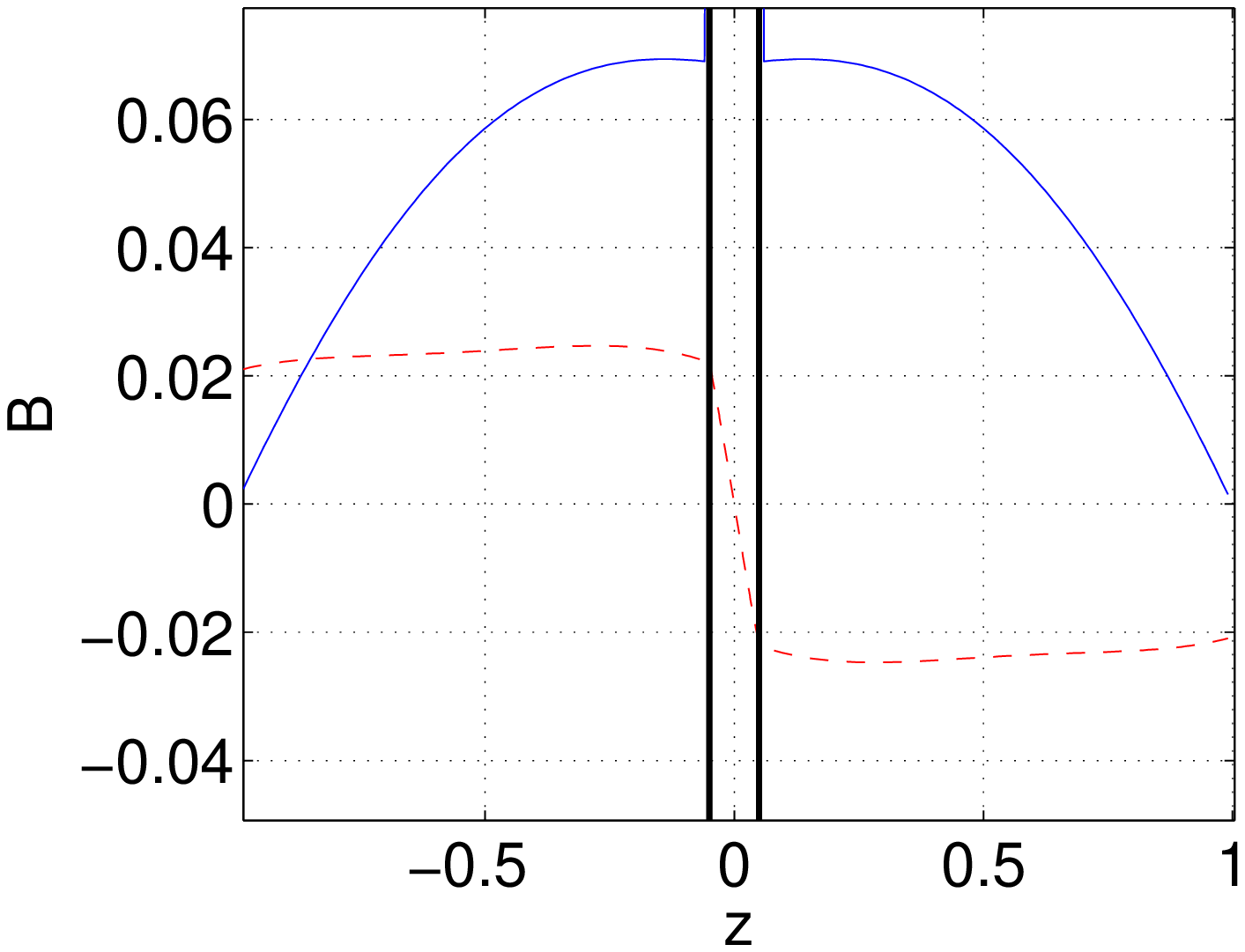}
\includegraphics[width=70mm,height=70mm]{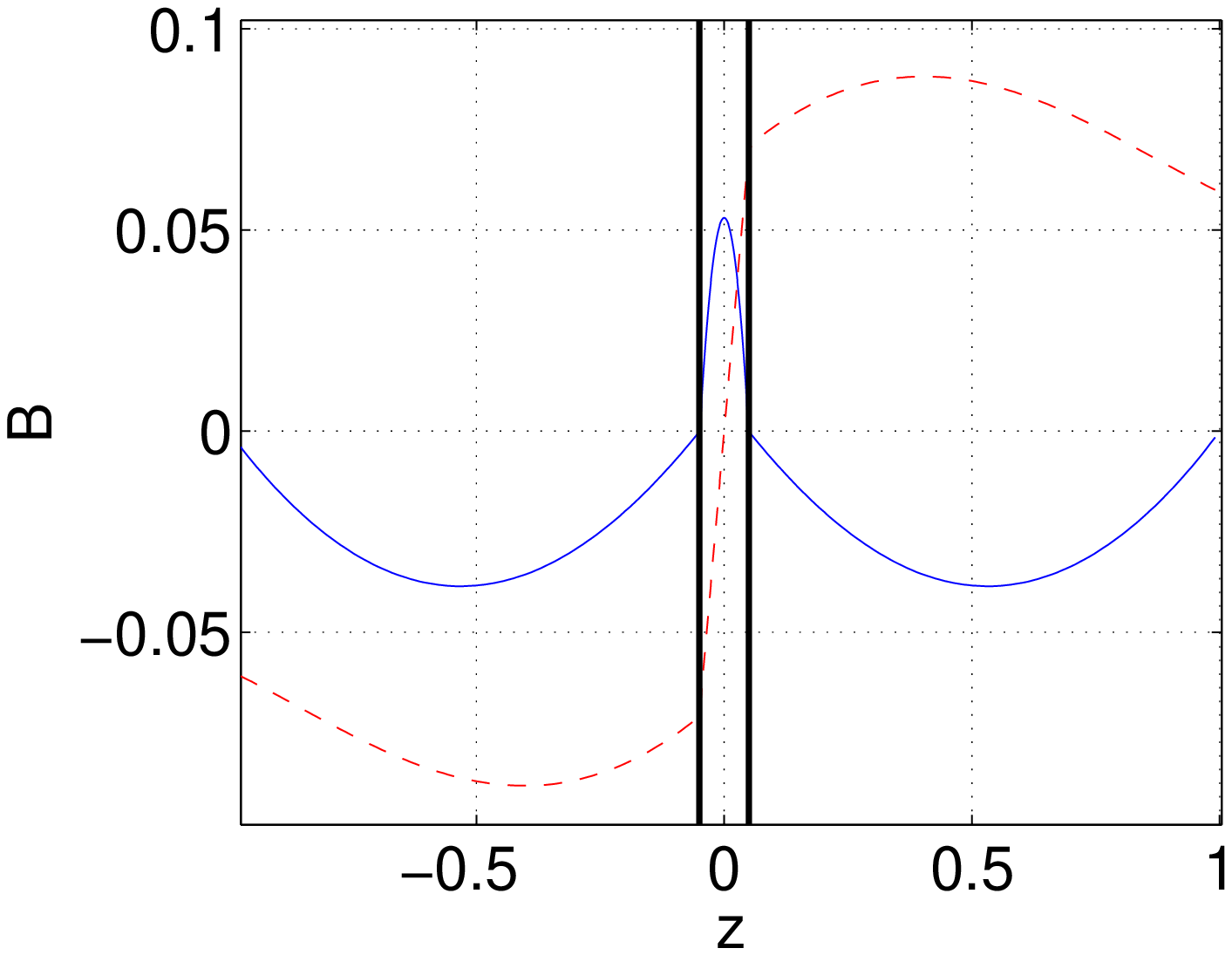}
\includegraphics[width=70mm,height=70mm]{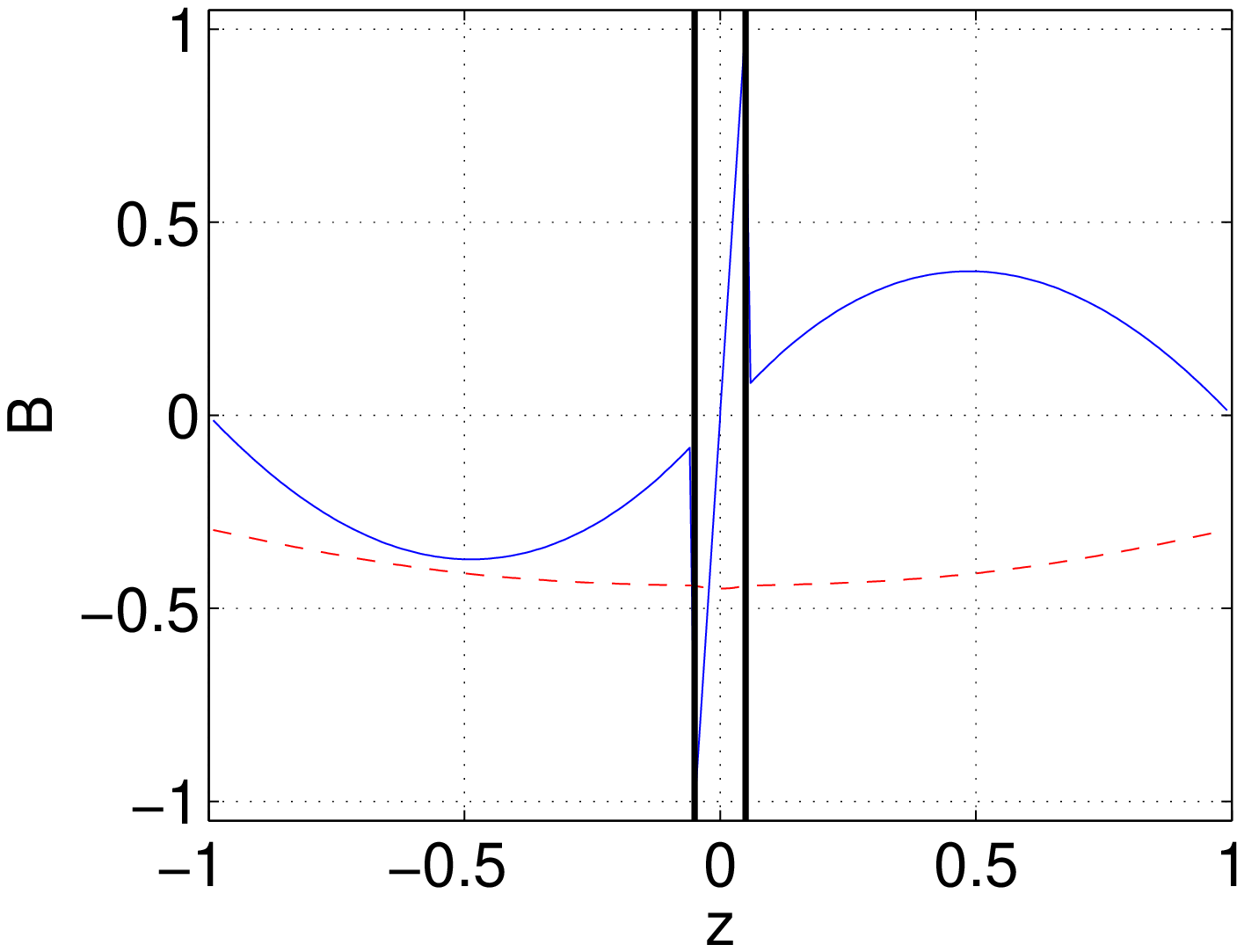}
\includegraphics[width=70mm,height=70mm]{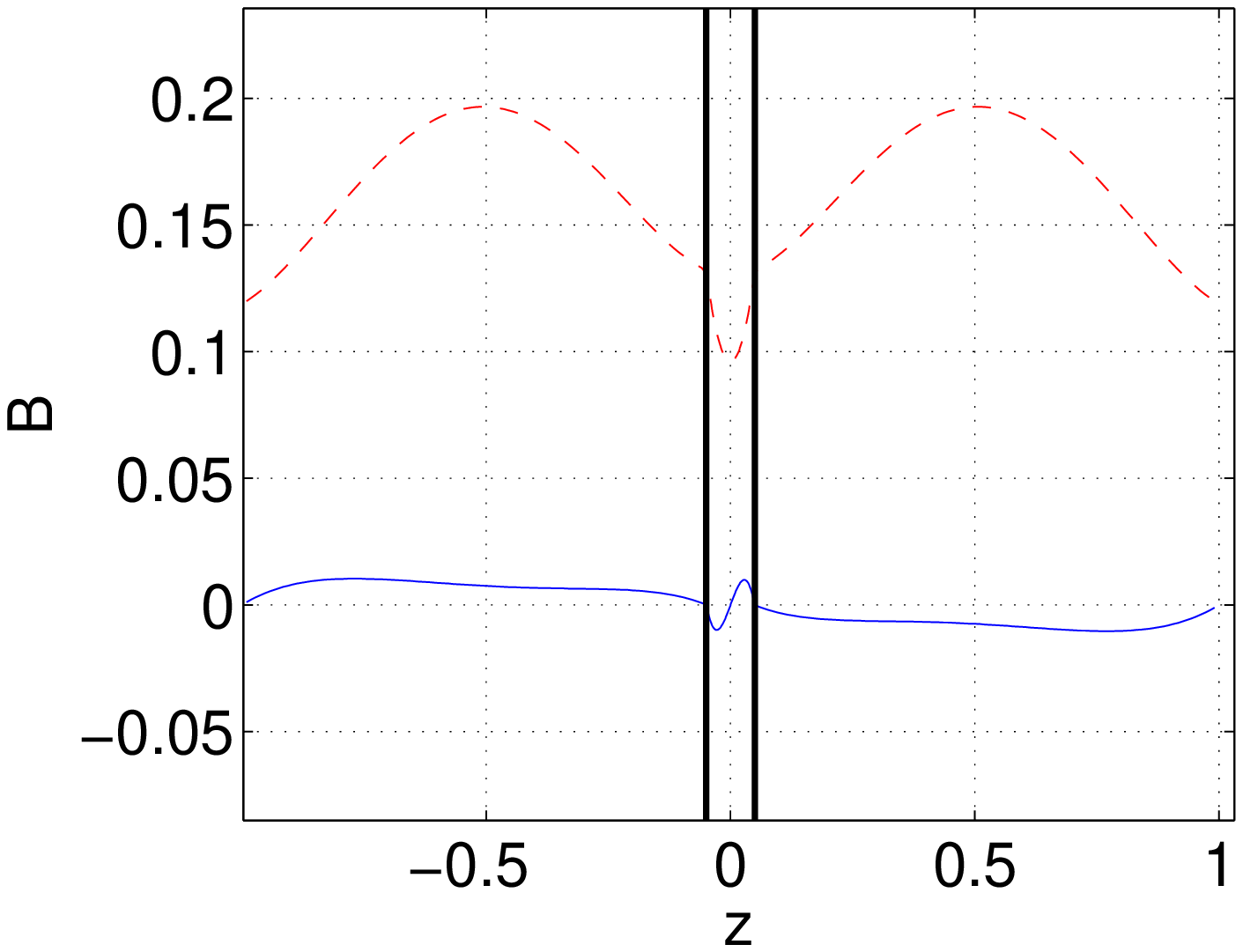}
\caption{Eigenmodes at their onset for a symmetric configuration and $\mu_r=14.5$, $\sigma_r=1$, top: most unstable mode  Rm$=6.475$. 
Bottom: second most unstable mode Rm$=8.17$. Real part (resp. imaginary part) are displayed on the left (resp. right). The $x$ (toroidal) component is the continuous (blue) curve. The $z$ (poloidal) component is the dashed (red) curve.  
}
\label{fig1b}
\end{center}
\end{figure}

\begin{figure}[htb!]
\begin{center}
\includegraphics[width=70mm,height=70mm]{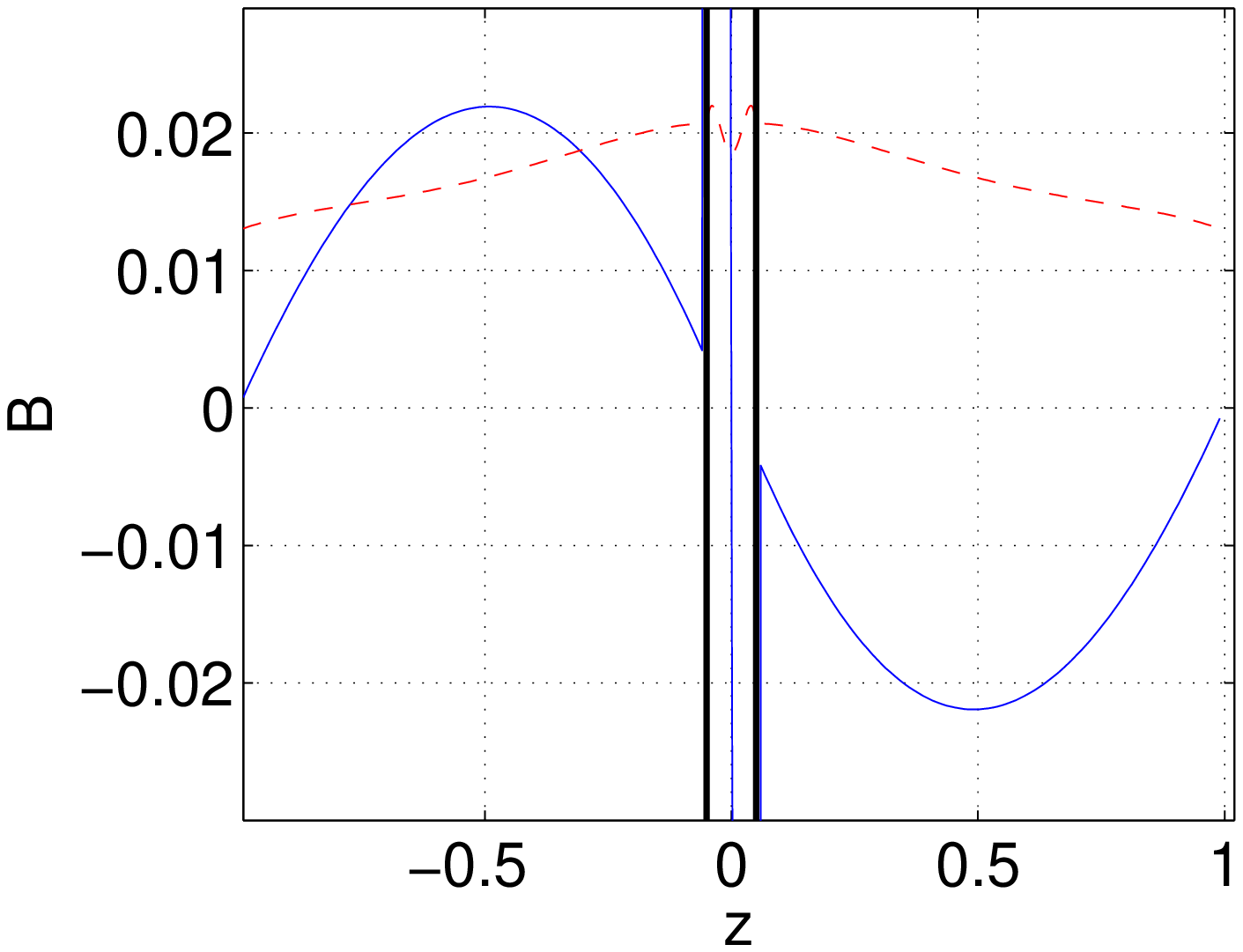}
\includegraphics[width=70mm,height=70mm]{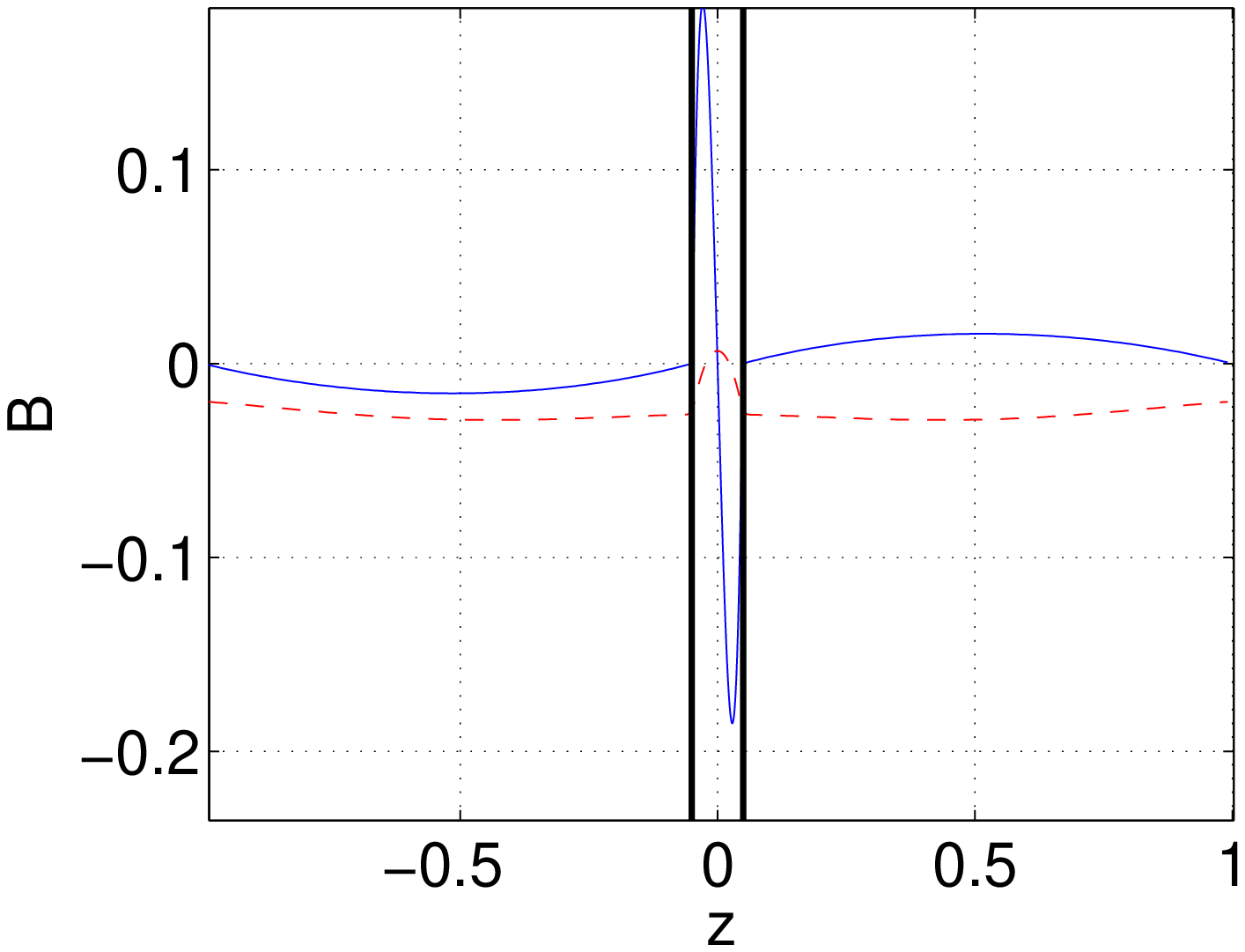}
\includegraphics[width=70mm,height=70mm]{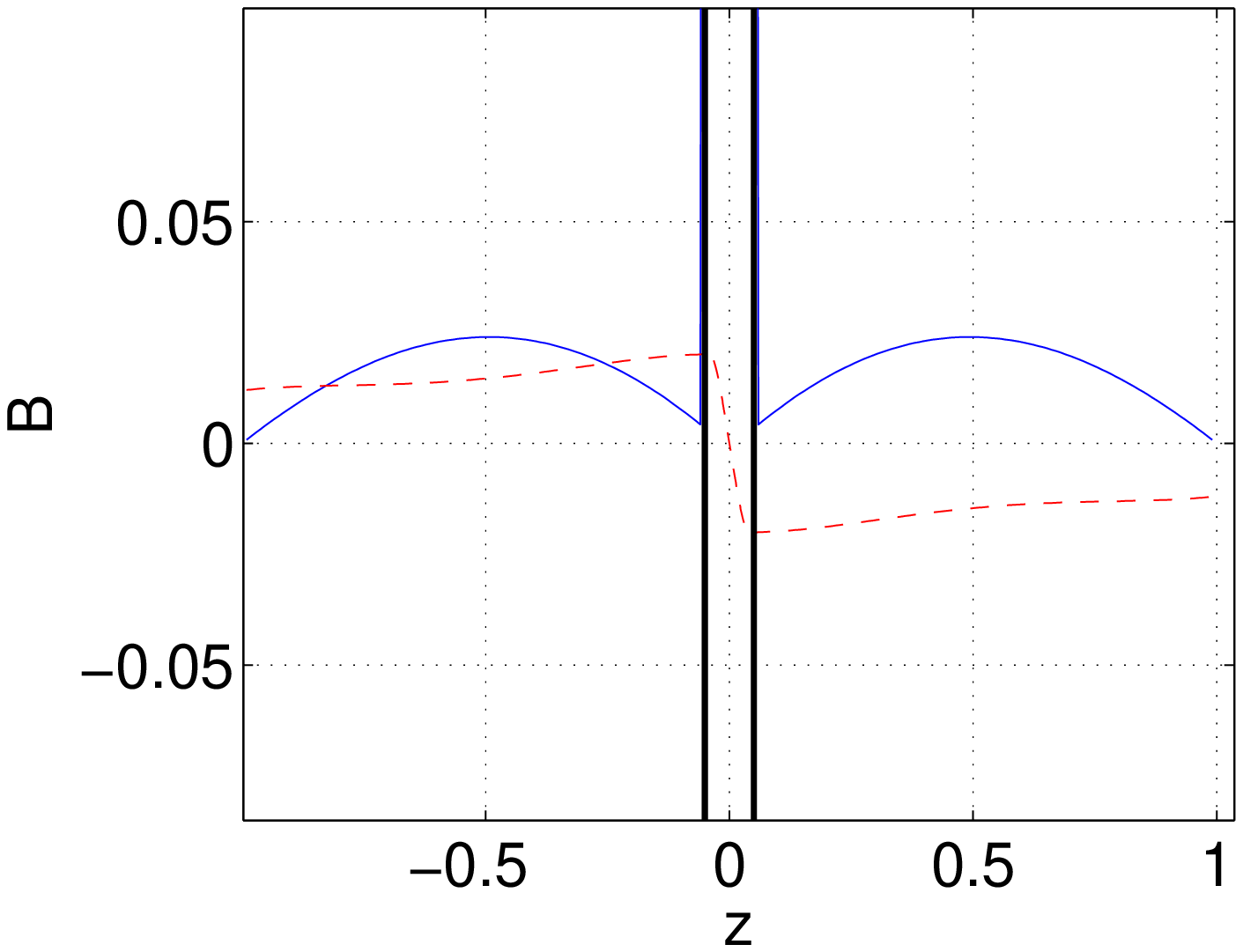}
\includegraphics[width=70mm,height=70mm]{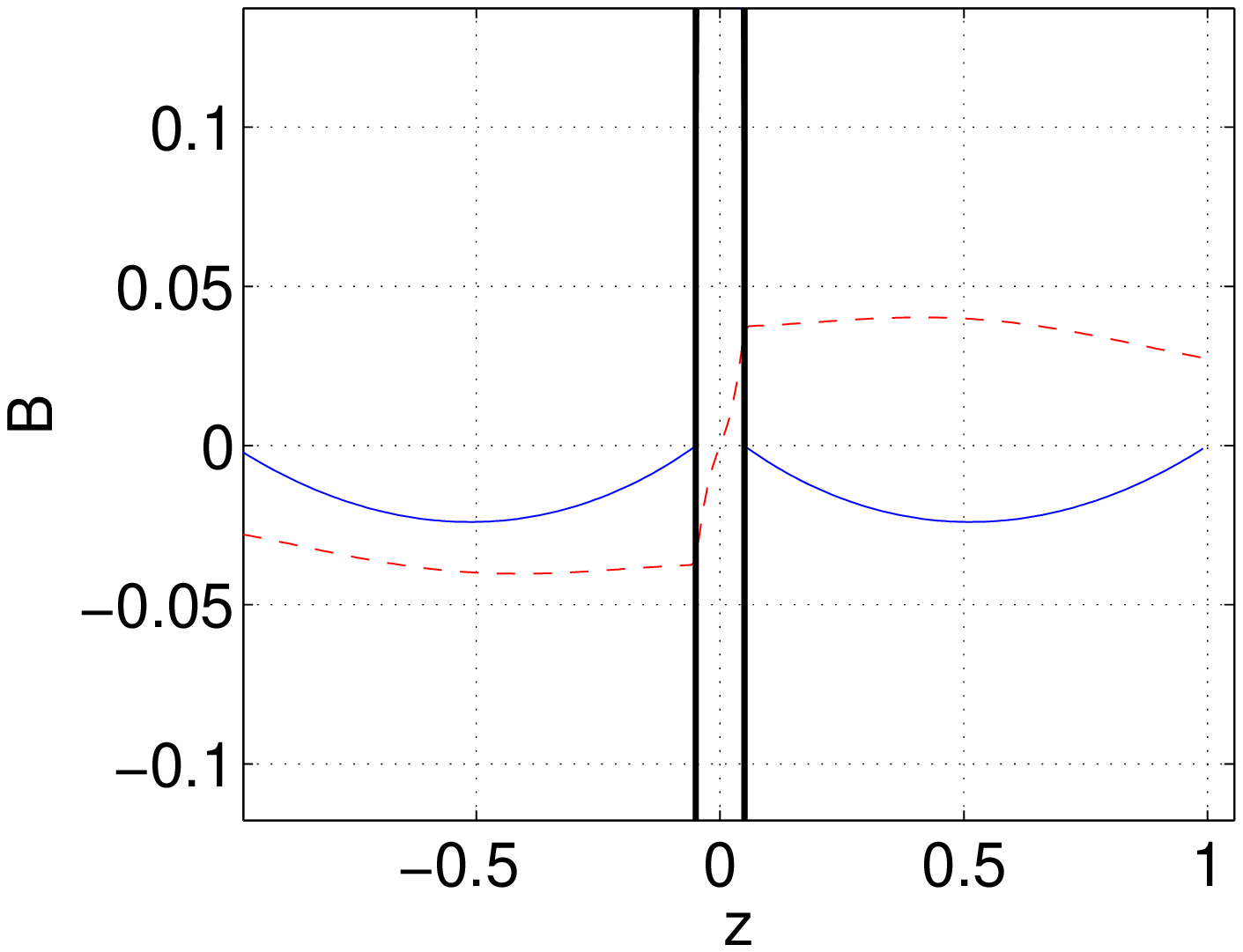}
\caption{Eigenmodes at their onset for a symmetric configuration and $\mu_r=300$, $\sigma_r=1$, top: most unstable mode  Rm$=8.36$;  
bottom: second most unstable mode Rm$=8.47$. Real part (resp. imaginary part) are displayed on the left (resp. right). The $x$ (toroidal) component is the continuous (blue) curve. The $z$ (poloidal) component is the dashed (red) curve.
}
\label{fig1c}
\end{center}
\end{figure}

The shape of the eigenmodes is presented in fig. \ref{fig1a}, \ref{fig1b},
\ref{fig1c}. The two modes have opposite symmetries. The continuous (blue) curve is the $x$ (toroidal) component
and the dotted (red) curve is the $y$ (poloidal) component. Since the
problem is invariant under $z-$reflection, the
eigenmodes are either odd or even under  this transformation.
Consequently, the components $b_z$ and $b_x$ are of  opposite parities.
For $\mu_r=1$, the even mode is the most unstable one but  both modes have comparable thresholds. This is also the case for very large $\mu_r$. As expected in
the limit of high permeability, the field is nearly  normal to the
mid-layer boundary at  $|z|=h$. For $\mu_r=14.5$, the odd mode is the most
unstable and  is responsible for  the decrease in the value of Rm at
onset. We observe that  even though the $z$ (poloidal)
component changes sign in the mid-layer, this is achieved
over the  short width  of the mid-layer, so that the field is
large everywhere in the fluid even close to the mid-layer.

\begin{figure}[htb!]
\begin{center}
\includegraphics[width=70mm,height=70mm]{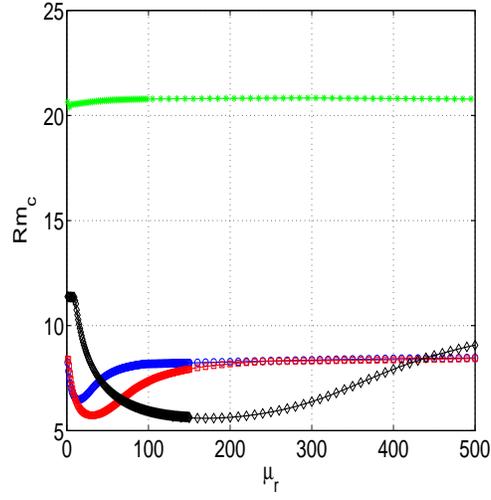}
\caption{Onset of the most unstable eigenmode  as a
function of the magnetic permeability $\mu_r$  of the mid-layer; for a
symmetric configuration $R_{\omega_1}=R\alpha_1=-R_{\omega_2}=-R\alpha_2=R_V=$Rm, $\sigma_r=1$, $h=1/20$ and (green star): $k=5$, (blue
circle): $k=1$, (red square): $k=0.5$, (black diamond): $k=0.1$.
}

\label{fig1tot}
\end{center}
\end{figure}

We have investigated how the transverse wavelength  affects  the reduction of the onset. The onset of the most unstable mode is presented
as a function of $\mu_r$ for different values of $k$ in fig.
\ref{fig1tot}.
For large $k$, the reduction of onset is very small. It increases when $k$
is decreased.
For instance, in the case $k=0.1$, the  onset is Rm$_c=11.3$
for $\mu_r=1$ and at very large $\mu_r$, is around Rm$_c\simeq 11$. The onset
is decreased to Rm$_c=5.6$ (a reduction around $50\%$) at $\mu_r=170$.

\begin{figure}[htb!]
\begin{center}
\includegraphics[width=70mm,height=70mm]{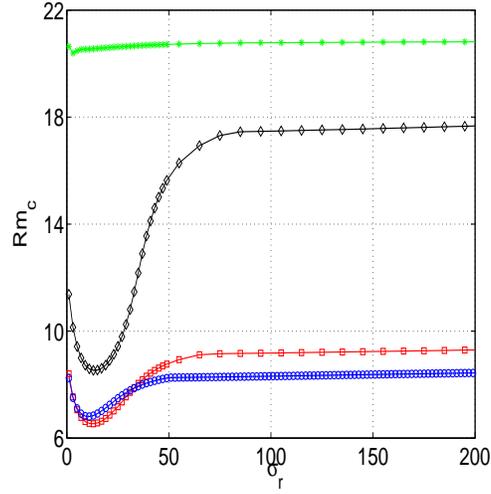}
\caption{Onset of the most unstable eigenmode  as a
function of the electrical conductivity $\sigma_r$  of the mid-layer; for a
symmetric configuration $R_{\omega_1}=R\alpha_1=-R_{\omega_2}=-R\alpha_2=R_V=$Rm, $\mu_r=1$, $h=1/20$ and (green star): $k=5$, (blue
circle): $k=1$, (red square): $k=0.5$, (black diamond): $k=0.1$.
}
\label{fig2}
\end{center}
\end{figure}

We have performed similar studies in the case $\mu_r=1$ and varying the
electrical conductivity of the mid-layer $\sigma_r$. Results qualitatively
similar to the case of varying $\mu_r$ are obtained for the onset
reduction, as displayed in fig. \ref{fig2}. However, the values of Rm$_c$ are quantitatively different from what is presented
in fig. \ref{fig1} so that there is no symmetry in exchange of magnetic
permeability and electrical conductivity, {\it i.e. }
Rm$_c(\mu_r,\sigma_r)\neq$ Rm$_c(\sigma_r,\mu_r)$, as expected from equations \ref{eqzBL1} and \ref{eqzBL2}.

The shape of the eigenmodes (not presented here) displays the following properties.  At large electrical conductivity, the field is nearly 
tangent to   the mid-layer boundary, as expected. At intermediate
conductivity, the smallest value of onset is achieved by a field whose $x$
(toroidal) component changes sign over a short length scale in the mid-layer
so that this component is large in the fluid even close to the mid-layer. We summarize  the effect of an increase of   conductivity by noting  that the  reduction of the value of Rm$_c$  is qualitatively similar to that  obtained for an increase of  the
magnetic permeability, but  the roles of the poloidal and toroidal components
are exchanged.

\begin{figure}[htb!]
\begin{center}
\includegraphics[width=70mm,height=70mm]{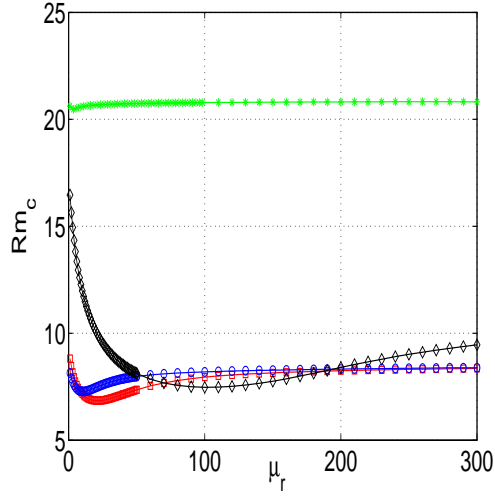}
\caption{Onset of the most unstable eigenmode  as a
function of the magnetic permeability $\mu_r$  of the mid-layer; for an
asymmetric configuration $R_{\omega_1}=R\alpha_1=-R_{\omega_2}=R_V=$Rm, $R\alpha_2=0$, $h=1/20$, $\sigma_r=1$ and (blue circle): $k=1$, (red
square): $k=0.5$, (black diamond): $k=0.1$.  An $\omega$ and an
$\alpha$-effect are operating in layer $1$ while only $\omega$-effect is
considered in layer $2$.
}
\label{fig5tot}
\end{center}
\end{figure}

In the VKS experiment, the role of the mid-layer is played by the disks.
These disks are fitted on one side with  blades. The two sides  of
the disks are thus different. If shear is present close to  both
sides,   helical vortices are expected to be more intense close to the
blades. A natural question is whether the effect that we have identified
in the case of a symmetric configuration (with induction  effects of
opposite intensities on both sides of the disks) remains efficient if the
flow is modified. We thus consider the case in which  an  $\alpha$-effect only
operates in one layer (layer $1$) and is zero in the other layer. We show the value of the threshold of the most unstable mode in fig.
\ref{fig5tot} for different values of $k$. We observe that a reduction
of the threshold also takes place in the asymmetric configuration. As in the
symmetric case, the effect is more intense when the wavelength
in the transverse direction is increased.

For the spatial structure of the modes, the system being  asymmetric,
the modes are neither odd nor even functions of $z$. Nevertheless,  some properties of the symmetric configuration are still verified:  compared to the case
$\mu_r=1$ or large $\mu_r$,   
the $z$ (poloidal) component of the most unstable mode displays a strong
gradient in the mid-layer for intermediate  $\mu_r$.
Consequently, the field is large everywhere in layer $1$ where both
induction effects are present.

It has been suggested that the disks in the VKS experiment are responsible
for an increased $\omega$-effect, the intensity of which increases with the
magnetic Reynolds number based on the disk magnetic permeability,
electrical conductivity, speed and size, {\it i.e.} $\mu_r \sigma_r R_V$ \cite{verhille}. To check whether a similar
description could explain the onset reduction that we have identified in
this work, we have varied $\mu_r$ keeping  fixed the  value $\mu_r \sigma_r=1$.  
A decrease in onset is observed at
intermediate values of $\mu_r$ which rules out  an effect
controlled by the  value of the magnetic Reynolds number of the layer.

Finally, since the disk electrical conductivity is different from that of the fluid in the liquid sodium experiment, we have changed the value of $\sigma_r$ and investigated how  Rm$_c$ depends on $\mu_r$. In the cases that we considered ($\sigma_r$ equals to $0.1$, $0.2$, $1$ and $5$), we always observe a decrease of the critical magnetic Reynolds number for a value of $\mu_r$ that slightly decreases with $\sigma_r$ (from $17.8$ for $\sigma_r=0.1$ 
 to $12.5$ for $\sigma_r=5$).

{\bf Description of the mechanism and conclusion}

We now describe the physical mechanism responsible for the decrease in onset.
For simplicity we consider the symmetric case.
The $\alpha$ and $\omega$ effects on either side of the layer have opposite
signs. This implies that the poloidal and toroidal components have the
same sign on one side and opposite signs on the other side.
Then, even though the $\alpha$ and $\omega$ effects change sign, in both domains
the induction mechanisms act to amplify the magnetic field.
As a consequence, one of the components vanishes at the center of the
mid-layer and
changes sign. In other words, from the point of views of symmetries, the magnetic modes must  be either odd or even with respect to reflection at the plane $z=0$,
which is what we have just obtained from the  discussion of the properties
of the
induction processes.

\begin{figure}[htb!]
\begin{center}
\includegraphics[width=70mm,height=70mm]{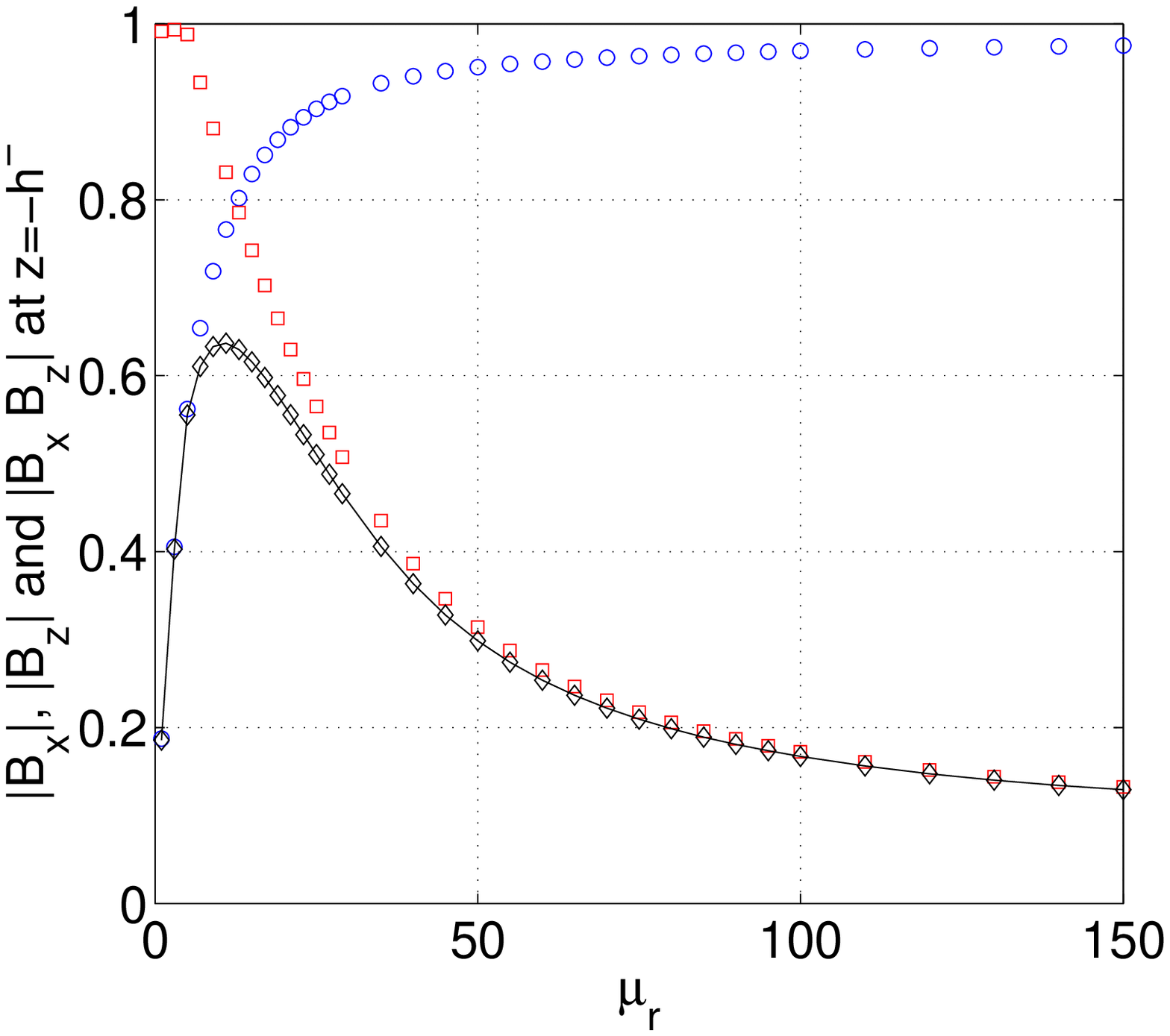}
\includegraphics[width=70mm,height=70mm]{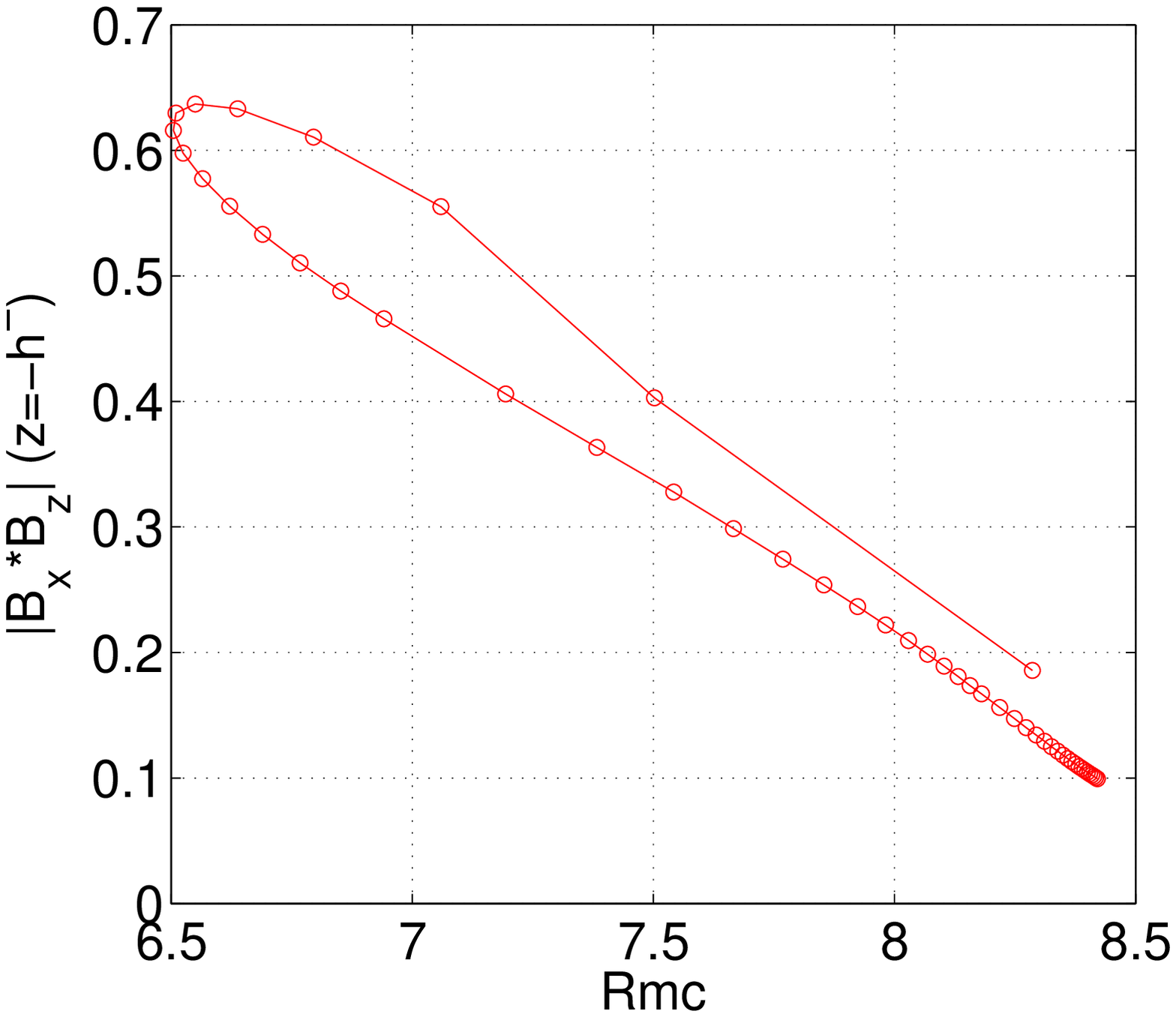}
\caption{Left: Value of $|B_x|$ (red square), $|B_z|$ (blue circle) and their product (black diamond) as a function of $\mu_r$ at the onset of the odd mode at $z=-h$ in the fluid. Same parameter values as in fig. \ref{fig1}. The fields are normalized by the maximum of $|B_x|$ and $|B_z|$ in the fluid. Right: $|B_x B_z|$ at $z=-h$ as a function of Rm$_c$ (same datas as left figure).}
\label{figend}
\end{center}
\end{figure}

For $\mu_r=1$, when one of the components changes sign in the mid-layer,  it is small
close to the mid-layer. There, the intensity of the induction
mechanisms is smaller than if the magnetic field were large. In the limit of infinite magnetic permeability, the tangential component
vanishes at the mid-layer boundary and the field is purely normal.  Just as for $\mu_r=1$, because the tangential component vanishes at the
mid-layer, the efficiency of one of the induction processes (the $\alpha$
effect) is also small near the mid-layer.  
A large but finite magnetic permeability can favor a dynamo mode that will
have a lower onset than in the $\mu_r=1$ or infinite $\mu_r$ limit.
Indeed, the normal component changes sign rapidly in the mid-layer and as
a consequence is large outside of the mid-layer. Thus, even though it changes
sign and vanishes in the center of the mid-layer,  at the fluid boundary the magnetic field is large
and the induction mechanisms act efficiently.
In other words, a large magnetic permeability allows the magnetic field to
change the sign of its poloidal component on a short length scale (while
the toroidal retains the same sign everywhere). The poloidal and toroidal
components are then both large  where the induction
mechanisms  operate. 
This allows for dynamo action at lower values of the magnetic Reynolds
number compared to the cases $\mu_r=1$ and $\mu_r=\infty$. To give some quantitative aspects to our argument,  we calculate $|B_x|$ and $|B_z|$ at $z=-h$ in the fluid for the odd mode. Results are shown in fig. \ref{figend}. In line with our discussion, $|B_z|$ (poloidal) increases  while $|B_x|$ (toroidal) decreases with $\mu_r$. Their product displays a maximum at a value of $\mu_r$ close to the value that minimizes Rm$_c$. More generally, the value of Rm$_c$ appears to be roughly a linear decreasing function of $|B_x B_z| (z=-h)$. Even though this holds only qualitatively  and is a consequence of the simple form of the eigenmode, it justifies our explanation for the effect of the mid-layer on the dynamo onset.

We have observed that the reduction of Rm$_c$   is always larger when the
wavelength in the transverse direction is large. This can be understood by
considering that this wavelength sets a possible length of variation in the
$z$ direction. If this length is short, then the field can change sign over 
a short length scale and gains a smaller benefit from an increase  of
$\mu_r$. In contrast, the lower the value of $k$, the larger the size
required by the field to change direction. Then an increase in magnetic
permeability is quite efficient in letting the field change sign over  a
shorter size.

\begin{figure}[htb!]
\begin{center}
\includegraphics[width=65mm,height=70mm]{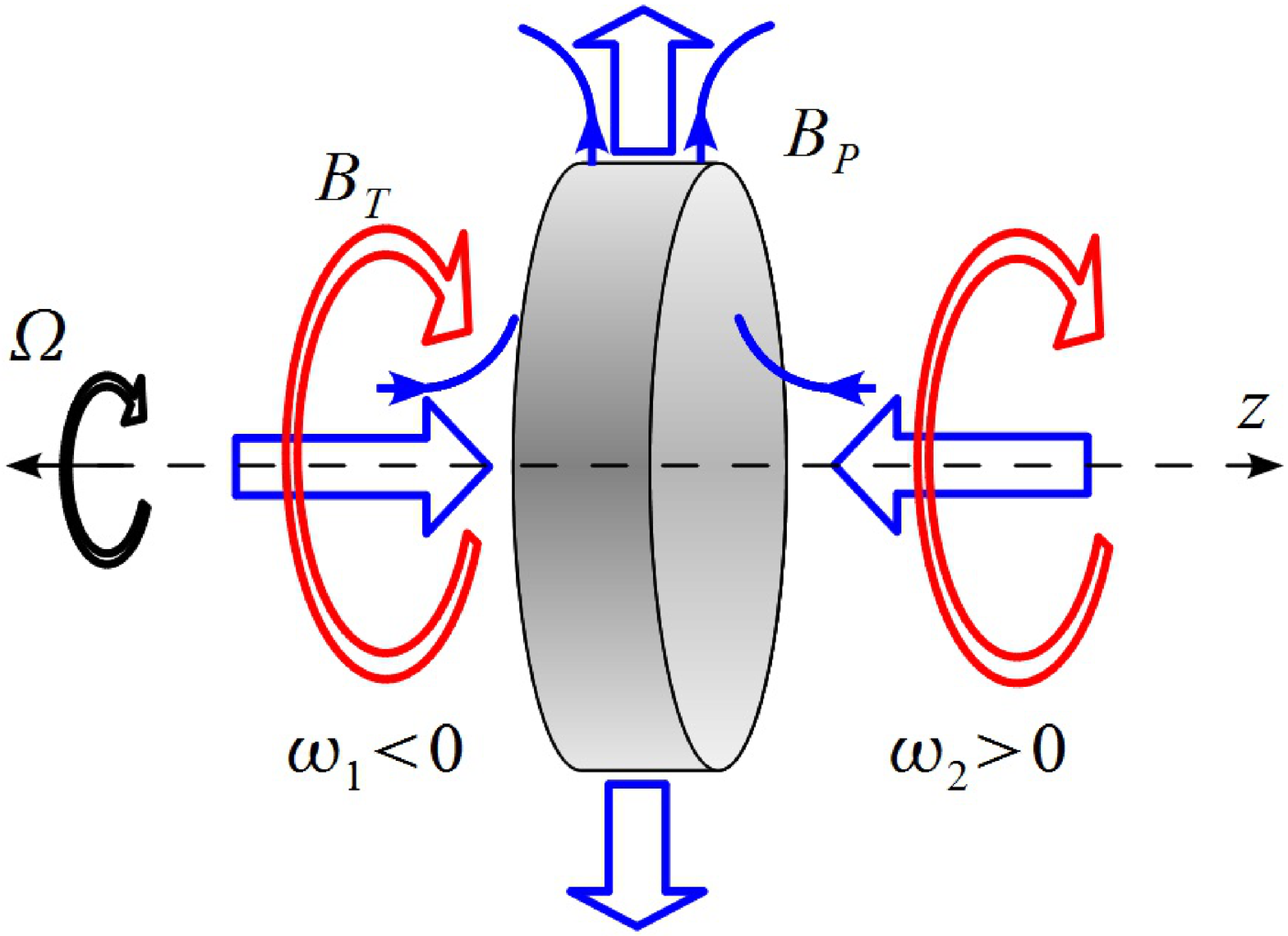}\includegraphics[width=75mm,height=80mm]{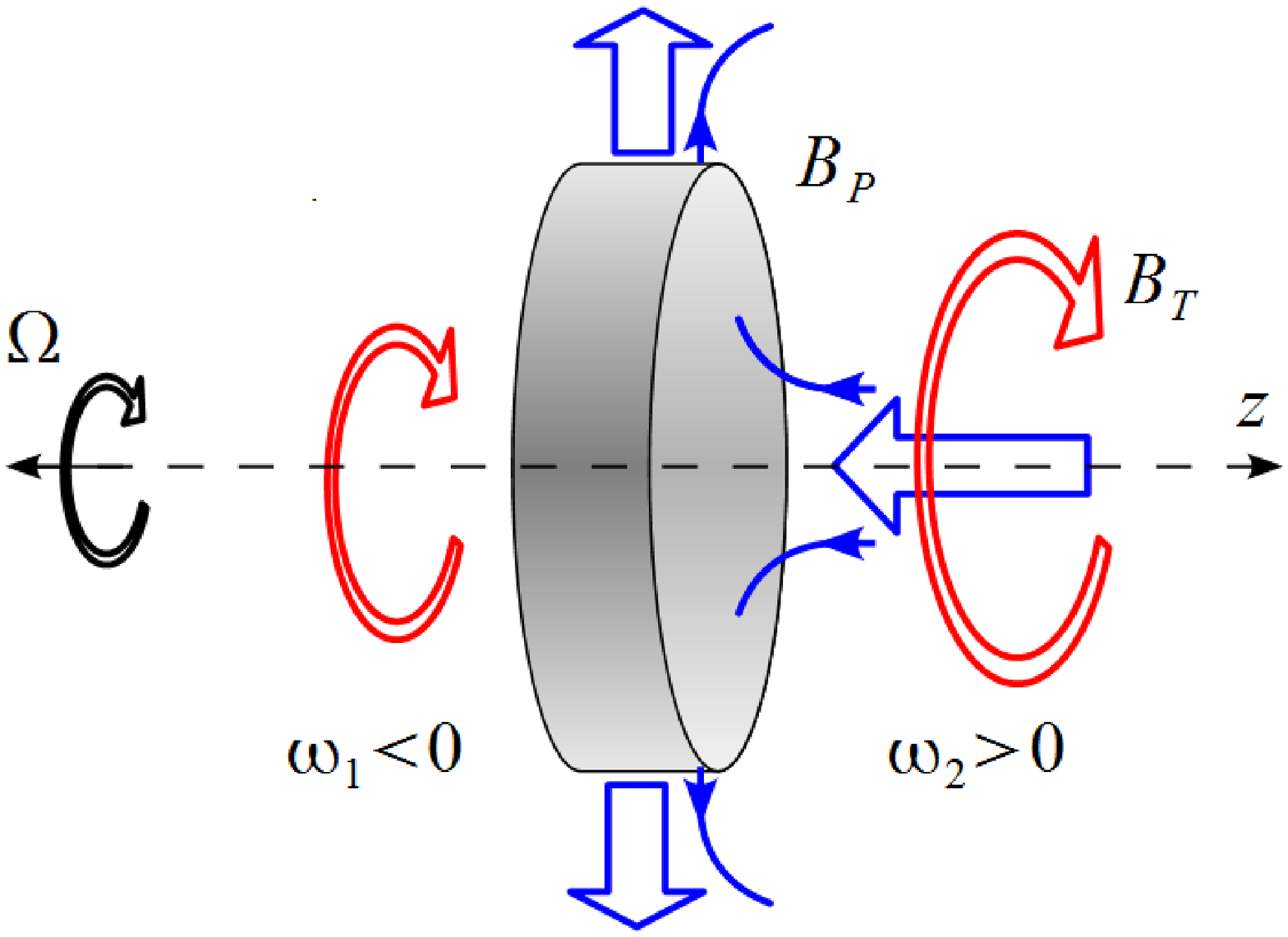}
\caption{Sketch of the large scales of the magnetic field generated by dynamo action  close to a rotating  disk (angular velocity $\Omega$) of large magnetic permeability. Left: symmetric configuration, the poloidal field (blue line) changes sign from one side to the other of the disk while the toroidal field is of same sign. The disk channels the magnetic field lines so that the poloidal field changes sign on a short distance and is large even close to the disk surface. Right: asymmetric configuration, in the case where the conversion of toroidal to poloidal is small on one side of the disk, the magnetic eigenmode is expected to have a toroidal component of same sign on both sides while the poloidal is dominant on one side of the disk.
}
\label{figlast}
\end{center}
\end{figure}

This mechanism gives a possible explanation for the observation of the VKS
dynamo when the disks are made of soft iron. In this case the spatial
structure of the large scale magnetic field close to the disk should have
the features presented in fig. \ref{figlast}. If induction effects were of equal and opposite  intensities on both sides of the disk, 
 the poloidal (resp. toroidal) field would have opposite (resp.
same) signs on the two sides. The poloidal component would then change sign in the
disk but would achieve a  large gradient in the disk and reach  large values
in the fluid close to the disk. As we mentioned, the actual geometry of the disks in VKS is asymmetric: one side is not fitted with blades and has a metallic cylinder in its center used to rotate the disk. We expect  that the toroidal component will be large on both sides while the poloidal component is channeled in the disk so that it remains large on one side of the disk despite its strong gradient in the disk.

We have not attempted to make quantitative  comparisons between  our 
numerical results and  the VKS experiment (although the chosen characteristic
lengths are reasonable).  Such quantitative agreement would be fortuitous
because  the induction mechanisms in the experiment are different from
uniform shear and $\alpha$ effects. However we have checked that the existence of a minimal value of Rm$_c$ for a value of $\mu_r$ larger than unity but finite is a robust phenomenon: it persists even if the induction mechanisms are asymmetric or if $\sigma_r$ is varied or if the aspect ratio is changed.  We expect  that a more realistic model
would display even larger reduction of the onset.
In particular if the sources are localized instead of uniform, our
mechanism suggests an enhancement of the onset reduction  when the sources
are close to the mid-layer. This is relevant for both an $\alpha$ effect due
to
swirling flows generated close to the blades \cite{gafd} and for a
conversion by the inhomogeneities of  the magnetic permeability of the blades \cite{gallet1}, \cite{gallet2}.



\end{document}